\documentclass[namedreferences]{kluwer}
\usepackage{klucite}
\usepackage{graphicx}

\def\kms{km\,s$^{-1}$}
\def\ngc{NGC7538-IRS1\,N\,}
\def\Mo     {\hbox{$\rm M_{\odot}$}}
\def\eq#1{\begin{equation} #1 \end{equation}}

\def\half   {\hbox{$\frac12$}}
\def\Dv     {\hbox{$\Delta v_{\rm D}$}}

\def\Ro     {\hbox{$R_{\rm o}$}}

\def\ro     {\hbox{$\rho_{\rm o}$}}

\def\Oo     {\hbox{$\Omega_{\rm o}$}}

\def\tauo   {\hbox{$\tau_0$}}

\def\ltsimeq{\,\raise 0.3 ex\hbox{$ < $}\kern -0.75 em
 \lower 0.7 ex\hbox{$\sim$}\,}
\def\gtsimeq{\,\raise 0.3 ex\hbox{$ > $}\kern -0.75 em
 \lower 0.7 ex\hbox{$\sim$}\,}

\begin{document}
\begin{article}
\begin{opening}
\title{NGC7538 IRS1 N:\\
       modeling a circumstellar maser disk}            

\author{M. \surname{Pestalozzi}\email{michele@oso.chalmers.se}}
\institute{Onsala Space Observatory, SE--43992 Onsala, Sweden}
\author{M. \surname{Elitzur}\email{moshe@uky.edu}} 
\institute{Dept. of Phys. \& Astr., Univ. of Kentucky,
           Lexington, KY 40506--0055, USA}
\author{J. \surname{Conway}\email{jconway@oso.chalmers.se}\\} 

\author{R. \surname{Booth}\email{roy@oso.chalmers.se}}
\institute{Onsala Space Observatory, SE--43992 Onsala, Sweden}

\begin{ao}
Kluwer Prepress Department\\
P.O. Box 990\\
3300 AZ Dordrecht\\
The Netherlands
\end{ao}

\begin{abstract} 

We present an edge-on Keplerian disk model to explain the main component of the
12.2 and  6.7\,GHz methanol maser emission detected toward \ngc. The brightness
distribution and spectrum of the line of bright masers are successfully modeled
with high amplification of background radio continuum emission along velocity
coherent paths through a maser disk. The bend seen in the
position-velocity diagram is a characteristic signature of differentially
rotating disks. For a central mass of 30\Mo, suggested by other observations,
our model fixes the masing disk to have inner and outer radii of $\sim270$\,AU
and $\sim750$\,AU.

\end{abstract}


\end{opening}

\section{Introduction: disks in high-mass star formation regions}

Disks are expected to form during protostellar collapse, and low-mass stars
seem to provide good observational evidence for the existence of disks (e.g.
\opencite{qi03}; \opencite{fue03}). The situation is less clear for high-mass
stars. While 
in several cases velocity gradients in massive star forming regions have been
detected on large scales ($>10,000$AU, e.g. \opencite{san03}), evidence for
compact disks on scales $\le 1000$\,AU remains sparse. One possible example is 
IRAS20126+4104, studied by \inlinecite{ces99}.

Class II methanol maser emission, a signpost of high mass star formation
\cite{min01}, offers a potential indicator of disks since it often shows
linear structures both in space and position -- velocity
diagrams (\opencite{nor98}; \opencite{min00a}). One of the most striking
examples of a maser line in both space and velocity is found in \ngc
(\opencite{min98}). We present here the first quantitative 
Keplerian disk analysis of this maser without invoking the assumption of a
single radius. We find compelling evidence for the disk interpretation in this
case.

\begin{figure}
\includegraphics[width=10cm]{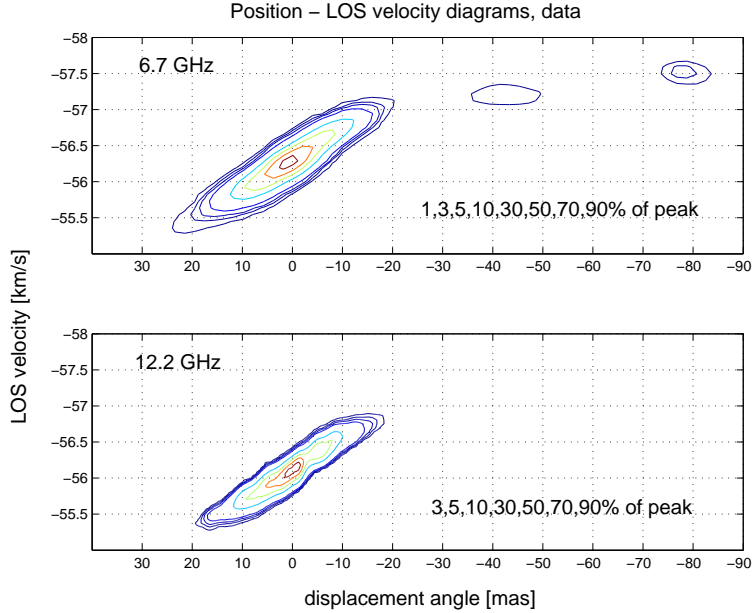}
\caption{Position -- velocity diagrams of the 6.7 (top) and
  12.2\,GHz (bottom) maser data of \ngc. The outliers visible in the 6.7\,GHz
     data were not considered for the fitting. }
\label{fig:ravel}
\end{figure}

\section{Basic theory of the model and fit to the data}

Because of the remarkable agreement of the positions of the maser emissions at
6.7 and 12.2\,GHz (2--3 mas), we draw the conclusion that the maser is the
result of the amplification of a background continuum. 
Consider than an edge-on rotating disk at a distance $D$ from the Sun and a
point at radius 
$\rho = r/D$ along a path with displacement $\theta$. The rotation velocity
$V(\rho)$ and its line-of-sight component $v$ obey $v/\theta = V/\rho =
\Omega$, the 
angular velocity. In Keplerian rotation $\Omega \propto \rho^{-3/2}$ with $\Oo
= \Omega(\ro) = D \sqrt{GM/\!R_{\rm o}^3}$ where $\ro = \Ro/D$ is the outer
radius and $M$ the central mass. Assume a Gaussian frequency profile for the
maser absorption coefficient and denote the radial variation of its magnitude
at line center by the normalized profile $\eta(\rho)$ ($\int\eta\,d\rho = 1$).
Then
\eq{\label{eq:tau}
 \tau(\theta,v) =  \tauo \int\!\eta(\rho)
      \exp\left[-\half\!\left(v -  \Omega(\rho)\,\theta\over\Dv\right)^{\!2}
      \right] {d\rho\over\beta}
}
where $\beta = \sqrt{1 - (\theta/\rho)^2}$ and $d\rho/\beta$ is distance along
the path. The width \Dv\ is taken as constant and we set it to be the 
spectral line width of the maser emission at 12.2\,GHz at $\theta = 0$. In
our case \Dv = 0.4 \kms. This value could vary with $\rho$ because of
temperature variation and maser saturation. Saturation 
typically requires $\tau \ge$ 10--15 across the disk radius, and the broadening
of the absorption profile is then proportional to length in excess of this
threshold (\opencite{eli92c}). The saturation issue is not treated in
this study. 

By comparing the maser brightness
temperature $T_b$ at $\theta = 0$ with $T_b$ of the background continuum (10000
-- 15000\,K, \opencite{cam84}) we find that $\tau_0 = 18.32$ and 15.99 for the
6.7 and 12.2\,GHz masers respectively. We parameterize the function
$\eta(\rho)$ as $\propto \rho^{(-p)}$ and set $p = -0.5$, being the value that
gives the best fits. The adopted distance to the source $D$
is 2.7\,kpc. 

{\it Fitting.} The free parameters in the fit are the angular velocity
$\Omega_m$ and the ratio between inner and outer radii, $h$. We define
$\Omega_m = 1/2 (\Omega_i + \Omega_o)$, i.e. the mean of angular velocities at
the inner and outer radii, as the angular velocity measured in the data
(see straight slope in Figure \ref{fig:results2}). Because of the natural 
{\it scalefreeness} of the model, we have to give an estimate for the scale of
the disk, e.g. the outer radius. We choose a value which corresponds to a
central mass of 
30\,\Mo, relying on previous estimates in \ngc (e.g. \opencite{cam84}). It is
to notice however that the model can produce equivalently good fits when this
estimate is not taken into account. The model shows a lower limit at disk sizes
corresponding to central objects of subsolar masses. It still performs 
well when setting the central object to 100\,\Mo, although then the outliers at
6.7\,GHz cannot be part of the disk.

We fit the position -- velocity diagram of the 12.2 and
6.7\,GHz methanol maser data observed using VLBI (the latter without outliers,
see Figure \ref{fig:ravel}). The 12.2\,GHz data fit is considered the {\it
  reference fit} because of the higher spatial resolution in the data.

{\it The kink.} The most important dynamical feature of the model is surely
the bend in angular velocity appearing at $\mid\theta_k\mid \approx 17$\,mas,
visible only in the position -- velocity data at
12.2\,GHz. This is naturally reproduced by our model and it can be explained
in the following way. When $\mid\theta\mid \le \theta_k$ the coherence path
participating to the maser extends 
over the entire disk, leading to the highest $\tau$ and so the brightest maser
emission. When increasing the displacement above $\theta_k$, the longest
coherence path will be centered at velocities closer to the outer edge of the
disk, making the increase in angular velocity with displacement angle (or
velocity gradient) less pronounced. This will produce a clear bend in the line
tracing the maximum optical depth $\tau$ at every value of $\theta$. The value
for $\theta_k$ gets smaller and the bend stronger the  
smaller \Dv\ or the smaller $h$. The presence of the bend is a unique signature
for a differentially rotating disk (e.g. a Keplerian disk as the one we model
here), since a solid-body rotation would show a linear velocity gradient. We
consider the independent observation of the bend in data and model to be the
most compelling evidence for a circumstellar rotating disk in \ngc.  

Because of the high requirements in sensitivity and spatial resolution, as
well as on the smooth distribution of the masing material in the disk, this
bend was not seen when modeling the H$_2$O maser disk in NGC4258
(\opencite{wat94}). It has been
detected in thermal line observations and modeled with the help of multiple
dynamical components (e.g. \opencite{bel02}). Our modeling confirms that there
is no need for multiple dynamical components when trying to fit the bend.

\section{Results}

\begin{figure}
\includegraphics[width=6cm]{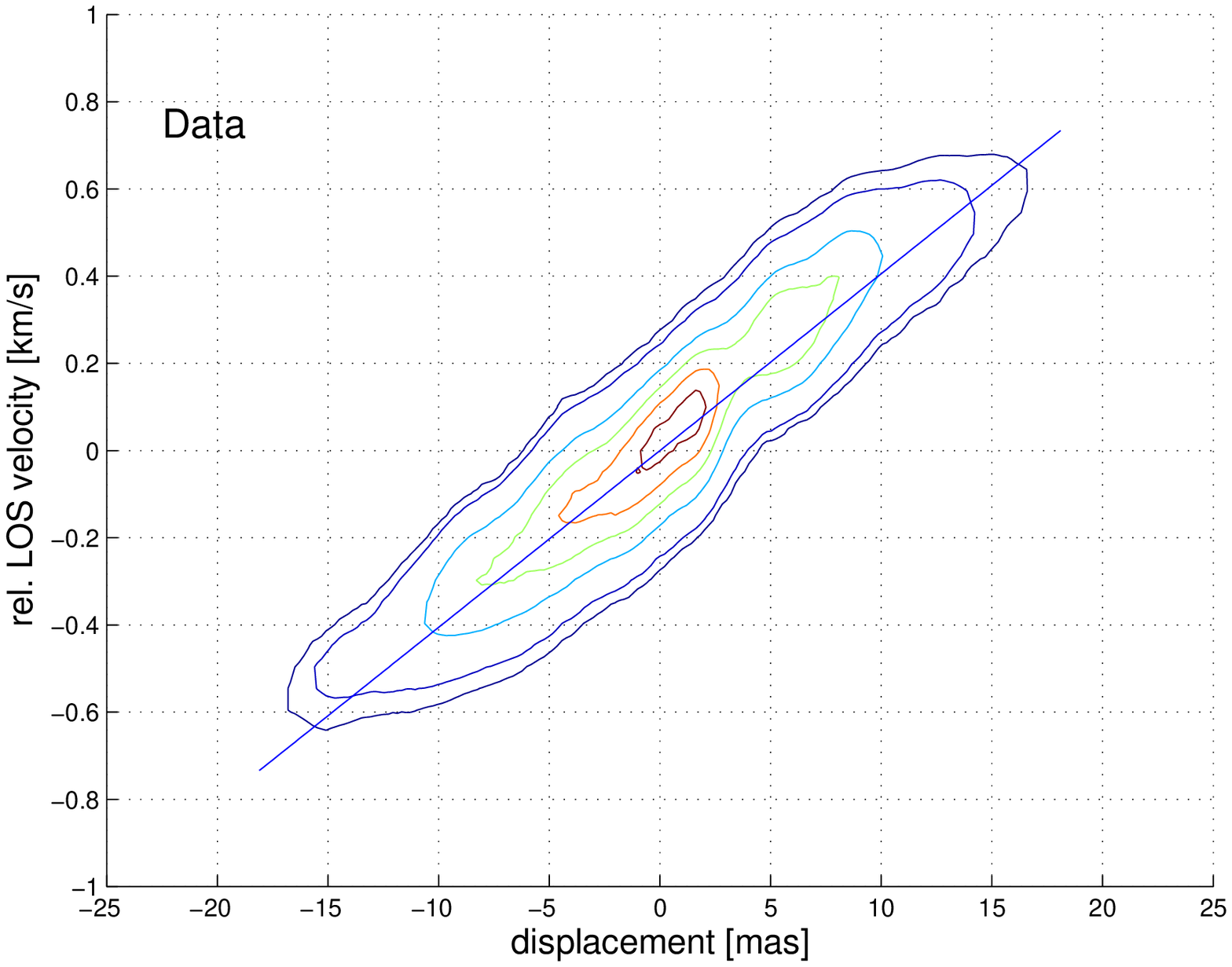}\includegraphics[width=5.7cm]{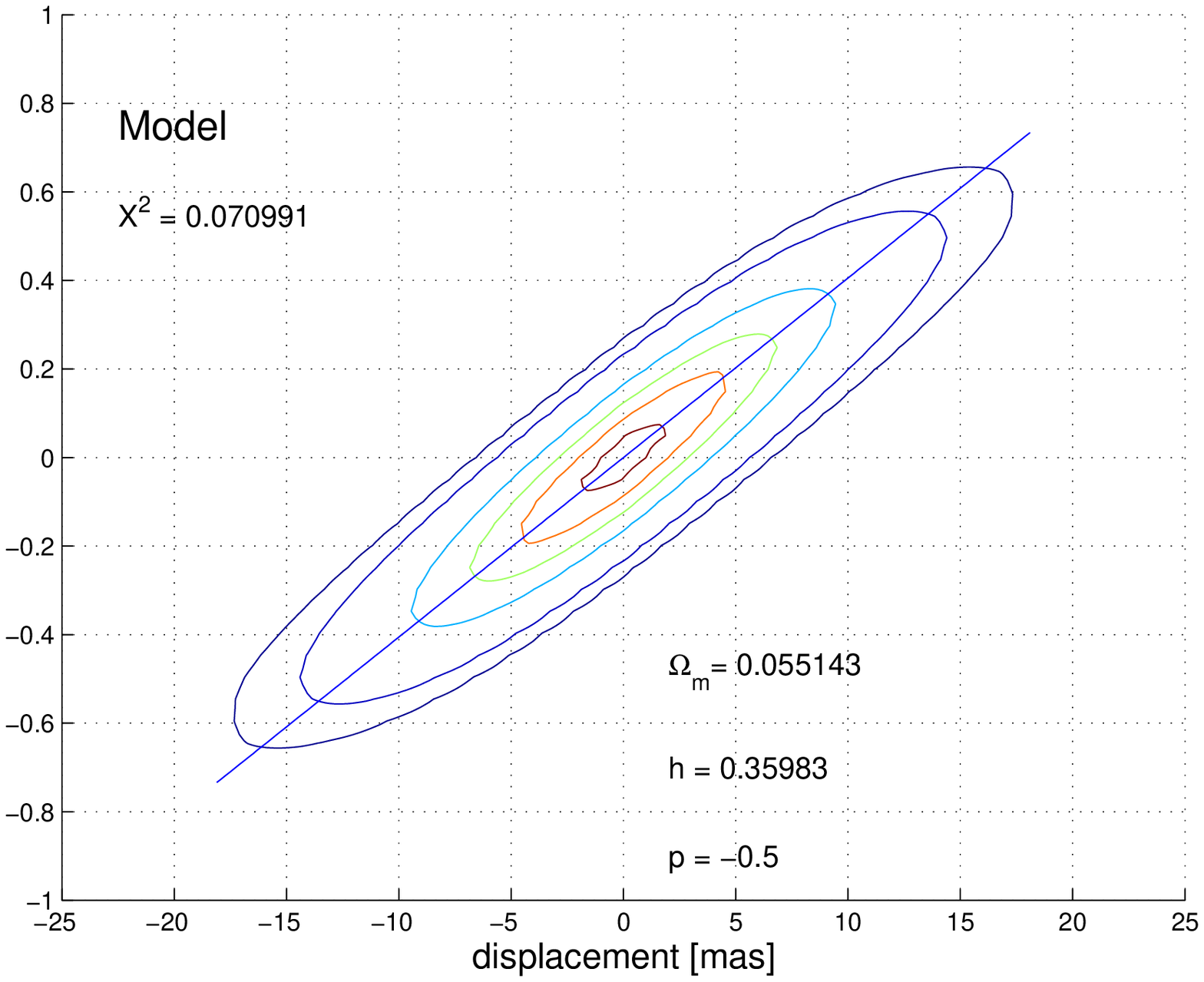}  
\caption{Data (left) and model (right) of the position --  
  velocity diagram of the optical depth $\tau$ of the main spectral feature of
  NGC7538 
  IRS1 N at 12.2\,GHz. In the right panel the values for $\Omega_m$ and $h$
  as well as an estimate of the goodness of the fit ($X^2$) are listed. The
  blue slope in both panels represents $\Omega_m$. } 
\label{fig:results1}
\end{figure}

\begin{figure}
\includegraphics[width=6cm]{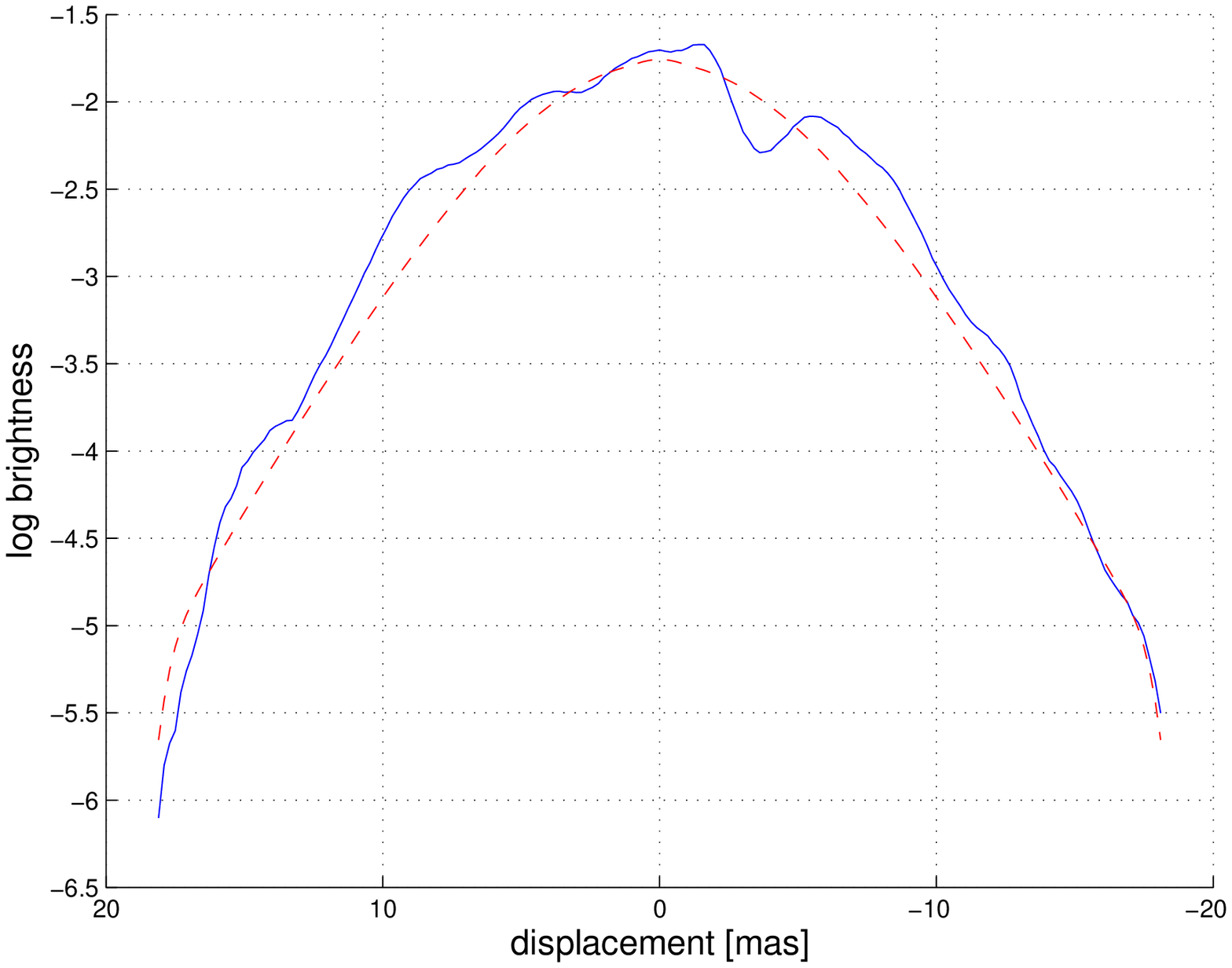}\includegraphics[width=5.8cm]{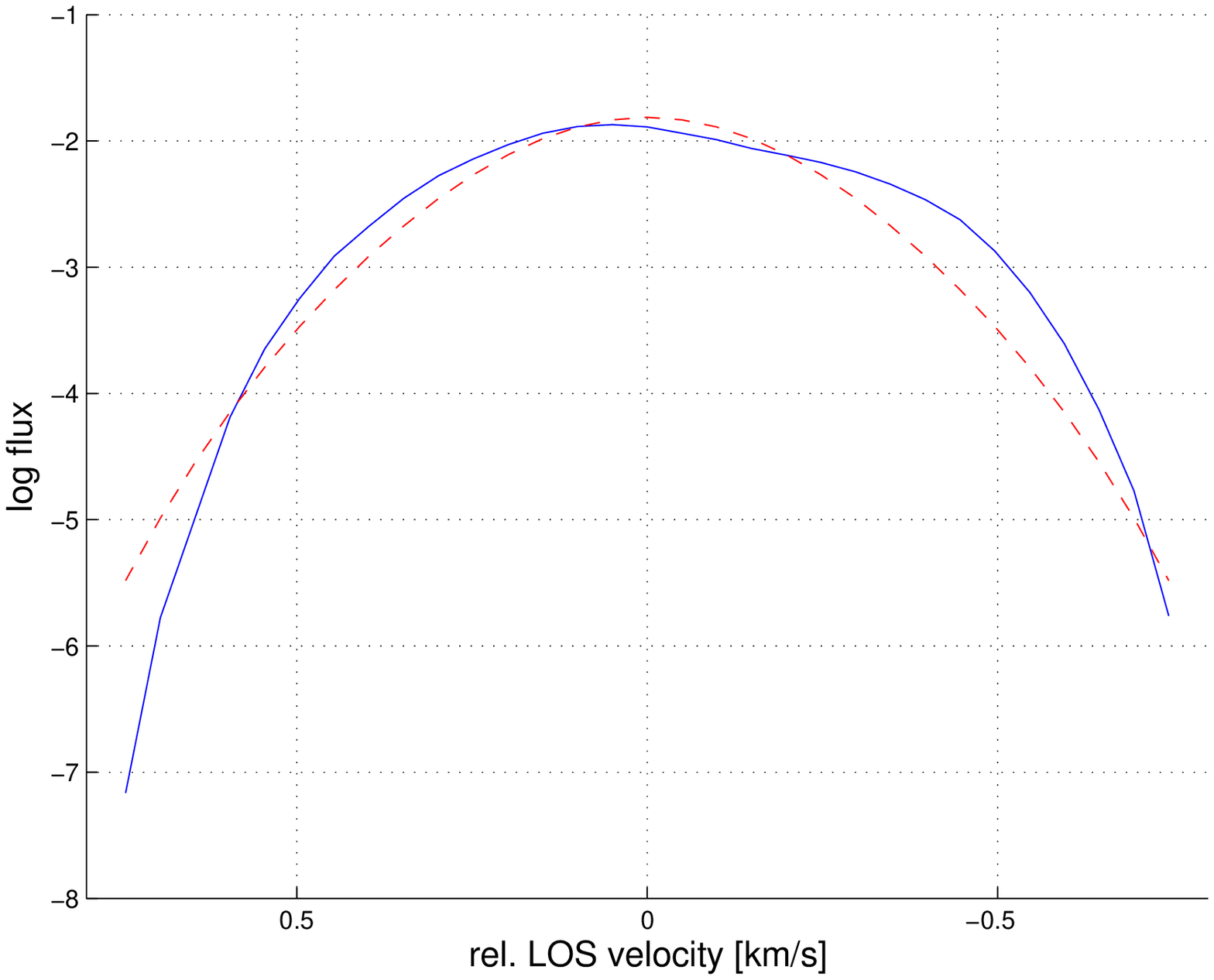} 
\caption{Brightness profile (left) and spectrum (right) of the main spectral
  feature of the methanol maser of NGC7538 IRS1 N at 12.2\,GHz. Filled blue
  lines are the data, dashed red lines the model.}
\label{fig:results2} 
\end{figure}

The results of the fitting of the 12.2\,GHz maser data is shown in the figures
\ref{fig:results1} and \ref{fig:results2}, $\Omega_m = 0.055$ and $h =
0.36$. These values correspond to a disk where the masing methanol is confined
between 270 and 750\,AU from the central protostar, which has a mass of
30\,\Mo. The model can reproduce with high accuracy both the  brightness
profile and the spectrum of the maser emission as well as the indications for
a bend in the velocity gradient. The fitting of the 6.7\,GHz data gave
similar results, i.e. $\Omega_m = 0.055$ and $h = 0.36$. Following the line of
maximum $\tau$ in the position -- velocity diagram in the
adopted model we can consider the outliers at 6.7\,GHz to be part of the disk.

\section{Final remarks}

The present study has produced the up to date most accurate model of a
circumstellar disk traced by maser emission at small scales ($ \ltsimeq
1000$\,AU) around a massive protostar. \ngc seems to be a special case, where
the masing material appears to be distributed extremely smoothly in the disk
and not to be particularly turbulent.  

The existence of the outliers in the 6.7\,GHz data however indicates that the
disk might have a more clumpy structure when approaching the outer edge: the
gravitational potential being weaker it is possible to imagine the disk to
get disrupted giving rise to local random enhancements of the optical depth
offering suitable conditions for maser amplification. In fact the outliers
seem to be very close to the line of highest optical depth in an edge-on disk
predicted by our model. 

A further issue is to decide whether the background continuum amplified by the
inverted methanol in the disk comes from the known UCHII region in \ngc and
illuminates the entire disk or from the central object solely. The latter case
seems to be in accordance with photoevaporation studies, where the radio
continuum would be responsible for both the methanol injection into the gas
phase and the detected radio continuum outflow (\opencite{hol94}). At the
moment it is not possible to discriminate both scenarios. 

Finally it is important to notice that the present study of maser emission
does not differ very much from more frequent studies of (optically thin)
thermal line emission. Once taken the logarithm of the maser brightness map the
problem reduces to the study of the value of the optical depth $\tau$, which
is exactly the same as in the case of thermal emission. The advantage of maser
emission is the possibility to achieve high spatial resolution and get
information about the dynamics at very small scales (e.g. 1\,AU at 1\,kpc). 

\acknowledgements
M. Pestalozzi thanks R. Parra for the Matlab tricks. We acknowledge
A. Jerkstrand for spotting a mistake in the original fitting program. The
12\,GHz data cube was kindly made available by V. Minier.

\bibliography{methanol,stars,othermasers,surveys_cat,tech,starformation+IR,varia,theory}
\bibliographystyle{klunamed}

\end{article}
\end{document}